\documentclass[draft,jgrga]{agu2001}
\usepackage{graphicx}
\authorrunninghead{YERMOLAEV ET AL.}

\titlerunninghead{COMMENT}

\authoraddr{Yu. I. Yermolaev, M. Yu. Yermolaev, I. G. Lodkina,
Space Plasma Physics Department, Space Research Institute, 
Profsoyuznaya 84/32, Moscow 117997, Russia. 
(yermol@iki.rssi.ru)} 

\begin{document}

%
%
%
%
%

%
%

\title{Comment on "A statistical comparison of solar wind sources of moderate 
and intense geomagnetic storms at solar minimum and maximum" by Zhang, J.-C., 
M. W. Liemohn, J. U. Kozyra, M. F. Thomsen, H. A. Elliott, and J. M. Weygand }
%

%
%



\author{Yu. I. Yermolaev, M. Yu. Yermolaev and I. G. Lodkina} 
\affil{Space Plasma Physics Department, Space Research Institute, 
Profsoyuznaya 84/32, Moscow 117997, Russia}

\begin{abstract}
(No abstract for Comment)
\end{abstract}

%
%

%

\begin{article}

%
%


Conditions in the solar wind resulting in magnetic storms on the Earth 
are a subject of long and intensive investigations. Recently 
\cite{Zh06}
[Zh06 hereafter],  published a paper, where they used superposed 
epoch analyses method to study solar wind features during 549 geomagnetic 
storms. Unfortunately, the used methodical approach has not allowed to 
improve essentially understanding of relation of magnetic storms with 
conditions in the solar wind, and first of all for the following reasons. 

1. Authors of Zh06 selected data on 4 different categories: 
(1) moderate storm at solar minimum, (2) moderate storm at solar maximum, 
(3) strong storm at solar minimum, and (4) strong storm at solar maximum. 
In the strict sense, this selection approach has not sufficiently serious 
physical arguments. 

On the one side, authors of Zh06 correctly noted that the storms are 
generated by different types of solar wind: ICME (MC) including Sheath 
and body of ICME and CIR (see, for instance, 
\cite{Vieira2004,HuttunenKoskinen2004,Yermolaev2005,YermolaevYermolaev2006}).
Used method leads to averaging corresponding parameters of different types 
of solar wind and, as result of this procedure, the calculated averaged 
parameters (for instance, density and temperature and parameters using 
them during calculations) are really not observed in the solar wind 
during magnetic storms.  As have been shown by 
\cite{YermolaevYermolaev2002} 
time variations in percentages of CIR-induced and ICME-induced storms 
have 2 maximuma (minimuma) per solar cycle and change with opposite 
phases. It means that result of averaging strongly depends, first of all, 
on real proportion between different types of solar wind included in  
selected time intervals, rather than on phase of solar cycle.

On the other hand, there is no experimental argument in favor of 
hypothesis that physics for moderate and strong storms may be different. 
Because strong storms are induced more often by ICMEs than CIRs, used 
selection of strong (moderate) storms results only in increasing 
(decreasing) portion of ICMEs in averaging database of solar wind. 
In this case result of averaging strongly depends on real proportion 
between different types of solar wind rather than level of Dst index 
using for storm selection.      

2. Authors of Zh06 took minimum Dst time as zero time for superposed 
epoch method. Because the main phase of storms may last from 2 up to 
15 hours (see, for instance, 
\cite{GonzalezEcher2005,Yermolaev2005,Yermolaevetal2006})
the shape of averaged Dst profile of storms significantly differ from 
shape of really observed storms and instead of onset (instant of storm 
start) has long (several hours) interval where parameters before and 
after onset have been averaged. In the strick sense, in contrast to 
usage of onset time as zero time (see for example 
\cite{LyatskyTan2003}
used method does not allow to select solar wind conditions before and 
after storm onset and to identify solar wind sources of storms. 

To illustrate mentioned above, Figs.1 and 2 present results of 
processing of OMNI data for 623 magnetic storms with Dst $<$ -60 nT 
during 1976-2000 
(\cite{YermolaevYermolaev2002,Yermolaevetal2006}):
time profile of averaged Bz (top panels), Dst and corrected Dst* 
parameters obtained by superposed epoch method with Dst storm onset 
and Dst minimum  as zero times, respectively. Fig.1. shows that the 
main phase of averaged storm lasts about 8 hours and time difference 
between minimum Bz and minimum  Dst is about 6 hours while in Fig.2 
(for the same zero time as in Zh06) there is no clearly defined main 
phase of the averaged storm and the time difference between minima 
Bz and Dst is only 1-2 hours, 
although in both cases the decrease in Dst index began in 1-2 hours 
after return of Bz component. The similar discrepancies are observed 
for several another time differences obtained with different zero times 
(see Figs. 3 and 4).

Differences in time profiles of solar wind and IMF parameters for CIR 
(121 storms), Sheath (22) and MC (113) are shown in Figs.3 and 4. 
We designated as "Unknown" also 367 storms for which there were not 
full set of measurements or the type could not be defined unambiguously. 
Figs. 3 and 4 use onset time and minimum Dst time as zero time, 
respectively, and show the same parameters: (Left column) N - density, 
V - velocity, Pdyn - dynamic pressure, T - proton temperature, T/Texp - 
ratio of measured proton temperature to calculated temperature using 
velocity, Dst index, (Right) $\beta$ - ratio of thermal to magnetic 
pressure, B, Bx, By and Bz - magnitude and GSM components of IMF and 
Kp index. Curves for different types of solar wind are presented by 
different color. The variability of data for all parameters and for 
all types of solar wind is sufficiently large, and therefore the 
table represents average values of their dispersion in the most 
disturbed and interesting part: from -12 till +12 hours relative 
to onset.  In several cases the distinctions between curves are 
less than corresponding dispersions, and in this case it is necessary 
to consider these distinctions as a tendency rather than a proved fact.

Because of place limit in the short comment we should discuss briefly 
the additional information arising owing to selection of data on solar 
wind types, and also advantages of choice of zero time. First of all 
one can see that the strongest storms were generated by sheaths but 
not by bodies of magnetic clouds. There are significant differences 
in T/Texp and $\beta$, for CIR and Sheath, on one hand, and MC, on the 
other hand. The highest value near onset is reached for density in 
CIR and for Pdyn in Sheath.  Detailed discussion and comparison of 
data in Fig.3 may be found in paper by Yermolaev et al., 2006. On the 
other side, comparison of Figs. 3 and 4 shows several advantages of 
choice of onset time as zero time. For example, Fig. 3 demonstrates 
that maximuma of density N and magnitude of magnetic field B for 
"Unknown" type and CIR are observed at storm onset and Fig.4 does not 
allow to make these conclusions.

%
\begin{table}
\caption{Anerage dispersions of solar wind and IMF parameters}  
\label{table:1}      
\centering                          
\begin{tabular}{l l l l l l l l l l l l l l}   
\hline 
   &&&&&&&&&& \\
SW & B & Bx & By & Bz & Tp & N & V & Kp & Dst & $\beta$ & T/Tex & NkT & $Nv^2$ \\
type& nT&nT&nT & nT & kK &cm$^{-3}$& km/s && nT &       &       &nPa  & nPa  \\ 
\hline
Unknown & 3.6& 5.2& 6.0& 4.6& 150& 8.1& 111& 13.1& 29& 0.57& 1.23& 0.033& 3.2\\
CIR     & 4.7& 6.7& 7.4& 6.2& 213&12.5& 102& 14.3& 32& 0.73& 1,51& 0.045& 4.2\\
Sheath  & 5.6& 5.2& 9.0& 7.1& 133&11.8&  88& 13.5& 36& 0.61& 1.00& 0.036& 7.7\\
MC      & 6.6& 7.1&11.0& 8.0& 138& 9.7& 128& 13.9& 37& 0.28& 0.87& 0.029& 5.5\\
\hline
\end{tabular} 
\end{table}
%

\begin{acknowledgements} 
       The authors are grateful to all databases for data used in the analysis. 
       Work was in part supported by RFBR, grant 04-02-16131, Program of Phys. 
       Dep. of RAS N 16.  
\end{acknowledgements}

\end{article}

\begin{figure}
\includegraphics[width=0.5\textwidth]{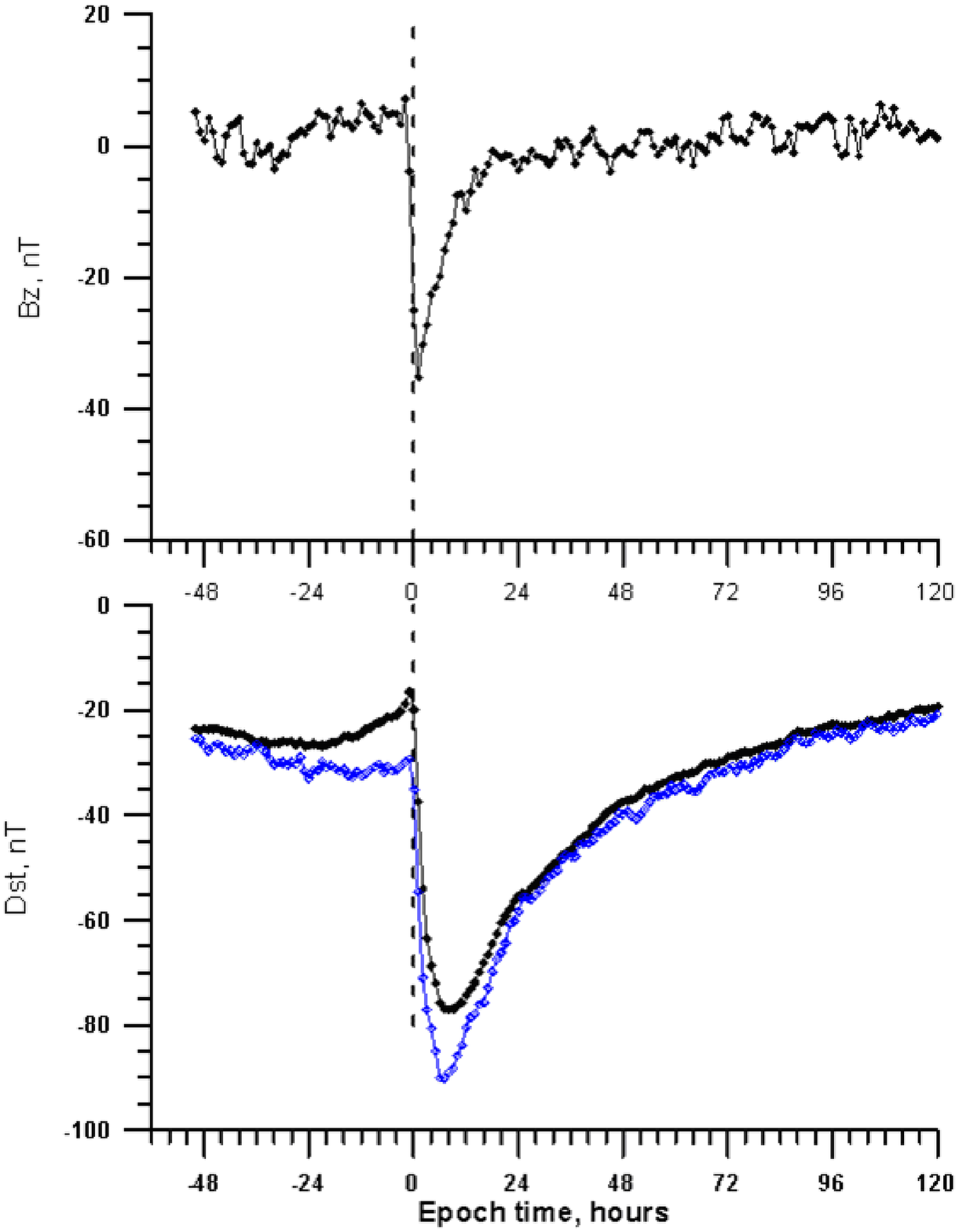}
\caption{Behavior of Bz IMF (top panel) and Dst (closed symbol) and 
corrected Dst* (open) indexes for 622 magnetic storms with Dst $<$ -60 nT 
during 1976-2000 obtained using OMNI dataset by superposed epoch method 
with zero time chosen as first 1-hour point of abrupt drop of Dst.} 
\end{figure}

\begin{figure}
\includegraphics[width=0.5\textwidth]{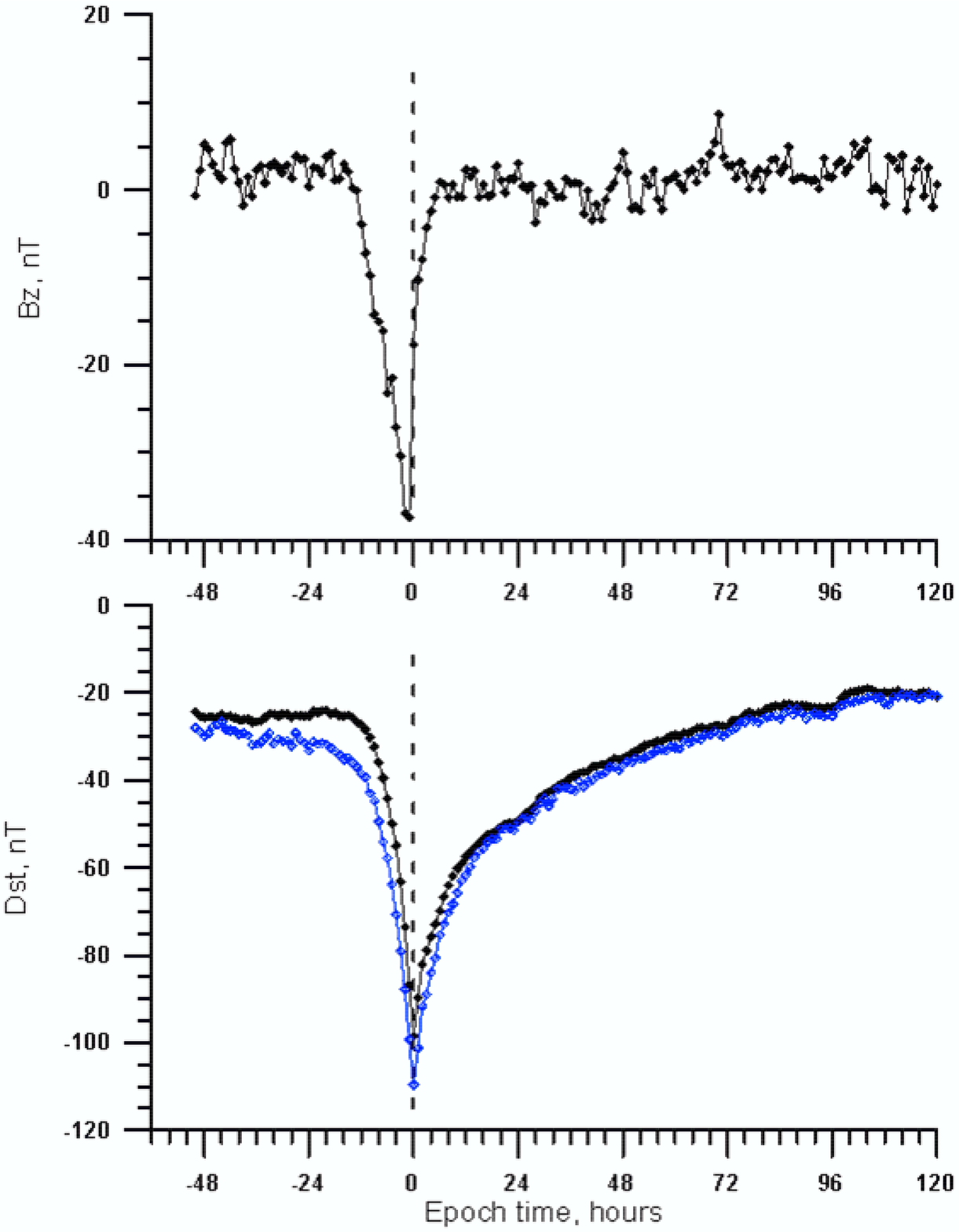}
\caption{The same as in Fig.1 obtained by superposed epoch method with 
zero time chosen as minimum of Dst (similar to Zh06).}
\end{figure}

\begin{figure}
\includegraphics[width=0.8\textwidth]{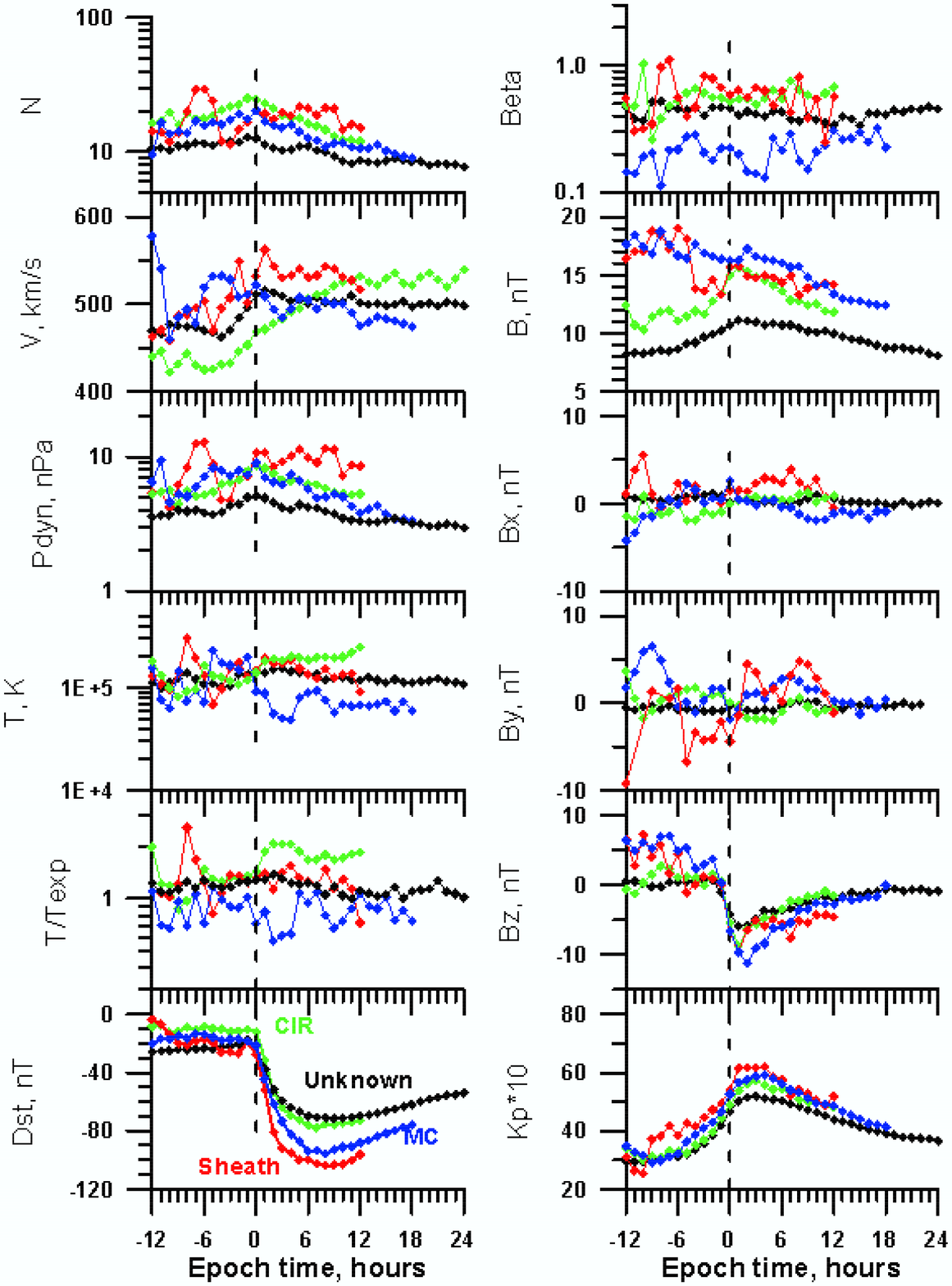}
\caption{Behavior of plasma and IMF for magnetic storms generated by CIR, 
Sheath, MC and Unknown types of solar wind during 1976-2000 obtained using 
OMNI dataset by superposed epoch method with zero time chosen as first 
1-hour point of abrupt drop of Dst.}
\end{figure}

\begin{figure}
\includegraphics[width=0.8\textwidth]{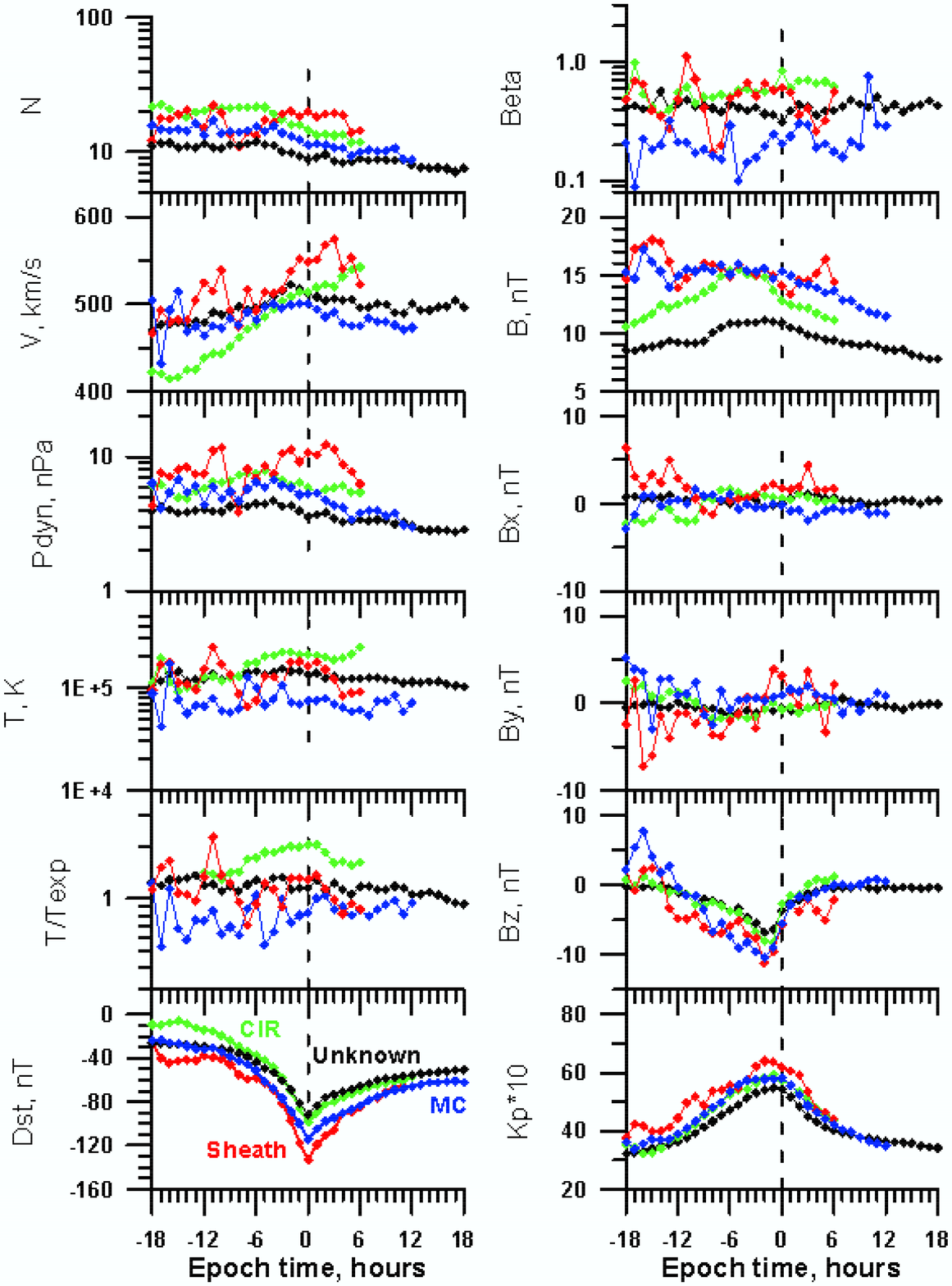}
\caption{The same as in Fig.3 obtained by superposed epoch method with zero 
time chosen as minimum of Dst (similar to Zh06).}
\end{figure}

\end{document}